\documentclass[10pt,conference]{IEEEtran}
\IEEEoverridecommandlockouts
\usepackage{cite}
\usepackage{amsmath,amssymb,amsfonts}
\usepackage{graphicx}
\usepackage{textcomp}
\usepackage{xcolor}
\usepackage{xspace}
\usepackage{float}  
\usepackage{dblfloatfix}

\def\BibTeX{{\rm B\kern-.05em{\sc i\kern-.025em b}\kern-.08em
    T\kern-.1667em\lower.7ex\hbox{E}\kern-.125emX}}

\newcommand{\sys}{\textsf{RevealNet}\xspace}

\usepackage{graphicx}
\usepackage{algorithm}
\usepackage{algpseudocode}
\usepackage{xcolor}
\usepackage{amsmath,amssymb,amsfonts}
\usepackage{cite}
\usepackage{booktabs}
\usepackage{tabularx}
\usepackage{caption}
\usepackage{subcaption}
\usepackage{paralist}
\usepackage{makecell} 
\usepackage{booktabs} 
\usepackage{xspace}
\usepackage{tikz}
\usepackage{multirow}
\usepackage{paralist}

\usepackage{array}
 \newcolumntype{L}[1]{>{\raggedright\arraybackslash}p{#1}}

\newcommand{\mypara}[1]{\vspace{1pt}\noindent{\bf {#1}}}
\newcommand{\myparait}[1]{\vspace{1pt}\noindent{\emph {#1}}}

\definecolor{figYellow}{RGB}{255, 215, 18}
\newcommand\encircle[1]{%
  \tikz[baseline=(char.base)] 
    \node[draw=none, align=center, shape=circle, inner sep=0.8pt, anchor=north, fill=figYellow, text=black]  (char)  {\footnotesize{\textbf{\textsf{#1}}}};%
}

\begin{document}

\title{\textsf{RevealNet}: Distributed Traffic Correlation for Attack Attribution on Programmable Networks\\
{}
}

\author{\IEEEauthorblockN{Gurjot Singh, Alim Dhanani, and Diogo Barradas}
\IEEEauthorblockA{\textit{David R. Cheriton School of Computer Science,} 
\textit{University of Waterloo}, Ontario, Canada \\
\texttt{\{gurjot.singh1,alim.dhanani,dbarrada\}@uwaterloo.ca}}
}
%

%
%
\IEEEoverridecommandlockouts
\IEEEpubid{\makebox[\columnwidth]{Manuscript accepted to the 23rd IEEE Intl. NCA Symposium.~\copyright2025 IEEE \hfill} \hspace{\columnsep}\makebox[\columnwidth]{ }}

\maketitle              
\begin{abstract}
Network attackers have increasingly resorted to proxy chains, VPNs, and anonymity networks to conceal their activities. 
To tackle this issue, past research has explored the use of traffic correlation techniques to perform \textit{attack attribution}, i.e., to identify an attacker's true network location.  
However, current traffic correlation approaches rely on well-provisioned and centralized systems that ingest flows from multiple network probes to compute correlation scores. Unfortunately, this makes correlation efforts scale poorly for large high-speed networks. 

In this paper, we propose \sys, a decentralized framework for attack attribution that orchestrates a fleet of P4-programmable switches to perform traffic correlation. 
We build on top of a set of correlation primitives inspired by prior work on computing and comparing flow sketches---compact summaries of flows' characteristics---to enable efficient, distributed, in-network traffic correlation. 
Our evaluation suggests that \sys achieves comparable accuracy to centralized attack attribution systems while significantly reducing the computational complexity and bandwidth overheads imposed by correlation tasks.

\end{abstract}

\section{Introduction}\label{intro}

In recent years, network attackers have increasingly relied on proxies, VPNs, and anonymity networks, to conceal their identities while engaging in malicious network activities. These anonymization tools route traffic through multiple intermediary servers, thereby obscuring an attacker's original IP address (i.e., a so-called \textit{stepping-stone} attack~\cite{zhang2000detecting}). Consequently, traditional approaches to identify the source of an attack, such as analyzing a flow's 5-tuple data, fail to trace malicious traffic effectively. This makes it challenging for network operators to perform \textit{attack attribution} (i.e., to locate the true source of attacks), preventing coordinated response efforts (e.g., via information sharing between ISPs), legal action, or better insights into attackers' tactics~\cite{untangling}.

To uncover the sources behind malicious and anonymized traffic~\cite{wang2002inter}, researchers have increasingly relied on traffic correlation techniques. These techniques aim to deanonymize malicious sources of traffic by analyzing and matching their traffic patterns (such as a flow's packets' timing and direction, and/or communication volume) as observed by multiple \textit{probe nodes} spread across the network~\cite{nasr2018deepcorr,lopes2024flow}. Previous studies developed statistical~\cite{murdoch2005low,palmieri2019distributed} and machine learning-based methods~\cite{nasr2018deepcorr,oh2022deepcoffea} to improve correlation accuracy. However, these methods require transmitting flows' features (as observed by the probes) to a central \textit{correlator node}, responsible for processing such features, thus leading to substantial network bandwidth and computational overheads. While decentralized approaches have been discussed~\cite{palmieri2019distributed,nasr2017compressive}, they mostly involve the partitioning of correlation tasks among multiple correlator nodes and still require probes to exchange flow features in bulk towards special-purpose servers, hence only partially mitigating scalability concerns (in particular, that of computation).

Addressing the scalability issues of existing attack attribution frameworks has proven particularly challenging in high-speed and large-traffic volume infrastructures (e.g., software-defined and programmable networks, such as those found in 5G deployments), where the volume of telemetry data grows rapidly with link speeds and which require rapid processing capabilities to uphold performance standards~\cite{sdn-traceback}.
To address constraints on data storage and the bandwidth overheads imposed by telemetry data offloading in the context of attack attribution, researchers have investigated the use of feature aggregation~\cite{lopes2024flow} and compression techniques that produce \textit{flow sketches}~\cite{coskun2009online,nasr2017compressive}, i.e., compact representations of flows' characteristics which can be used for correlation. 
While a significant step forward, we argue that sketches alone do not fully address the fundamental scalability limitations of attack attribution workloads.

This paper introduces \sys, a framework for attack attribution that operates via the decentralized correlation of attacking flows. \sys eschews the need for special-purpose correlation nodes and minimizes data exchanges during correlation tasks. At the core of our approach is the realization that, while flow correlation capabilities remain largely unexplored in P4-programmable switches (e.g., Intel Tofino, AMD Pensando), these devices are gaining traction in high-performance networks due to their ability to perform complex network security operations with low computational overhead~\cite{zheng2023network}. This raises the question of whether P4 switches can also leverage efficient correlation-focused flow feature extraction primitives---such as flow sketches---to operate as decentralized probe/correlation nodes, without incurring the additional costs of middlebox infrastructures~\cite{netbricks} or of offloading feature processing to dedicated servers~\cite{scalingHardware}.

Our evaluation suggests that \sys\ matches the effectiveness of centralized attack attribution systems while offering significant efficiency gains by decentralizing flow correlation. \sys allows P4 switches to track more flows and cut communication overheads---saving up to 96\% bandwidth in a decentralized setup consisting of 20 networks, each with a \sys-enabled switch.

\mypara{Contributions.} We summarize our contributions as follows:

\begin{compactitem}
    \item We design \sys, a decentralized attack attribution framework based on the orchestration of P4-programmable switches for enabling flow correlation.
    \item We implement \sys in \texttt{bmv2}, the reference P4 switch, and adapt prominent flow sketching schemes to fit the programming constraints of P4 switches.
    \item We evaluate \sys's correlation accuracy as well as its computational and bandwidth overheads when identifying the source of malicious flows.
\end{compactitem}

\section{Background and Related Work}

\subsection{Traffic Correlation}
\label{subsec:rw-correlation}

Traffic correlation techniques can be used to analyze traffic patterns and link together flows which are observed at the entry and exit nodes of proxy chains. While some correlation schemes helped gauge the privacy provided by anonymity~\cite{murdoch2005low} and mix networks~\cite{mix-match}, others were developed with the intent to trace stepping-stone attackers~\cite{holdingAccount,coskun2009online}. 
Below, we describe two main classes of prominent \textit{passive flow correlation} techniques: a) those that use fine-grained per-packet data for higher accuracy at the cost of increased storage, and; b) those that rely on coarse-grained per-flow data, which are more storage-efficient but are typically less precise.

\mypara{Flow correlation with fine-grained information.} Most studied traffic correlation techniques rely on fine-grained, per-packet information. 
Zhu et al.~\cite{zhu2009correlation} leverage per-packet timing information to compute the average traffic rate of flows at different intervals, while Palmieri~\cite{palmieri2019distributed} used wavelet-based analysis to capture timing, size, and rate variations across flows. 
Recently, researchers adopted deep learning to improve flow correlation, pushing accuracy over that of statistical methods. 
DeepCorr~\cite{nasr2018deepcorr} and DeepCoFFEA~\cite{oh2022deepcoffea} progressively improve accuracy---DeepCorr uses convolutional neural networks (CNNs) to learn correlation functions, and DeepCoFFEA introduces novel feature embedding and voting mechanisms.

While the above approaches yield high accuracy, they rely on the collection, communication, and processing of fine-grained information (direction, size, and timing) about packets in a trace, making them costly to deploy at scale (\S\ref{sec:attribution},\cite{nasr2017compressive}).

\mypara{Flow correlation with coarse-grained information.}
    Collecting and storing fine-grained traffic features for flow correlation at choke points (e.g., ISP border routers) is increasingly challenging due to the high volume and speed of traffic, which strain storage and processing resources. To overcome this, 
    Coskun et al.~\cite{coskun2009online} used linear projections to reduce a flow's packets' timing patterns into succinct representations that can be efficiently collected, stored, and compared. 
    Nasr et al.~\cite{nasr2017compressive} introduced compressive traffic analysis, a paradigm which leverages compressed sensing to compress the traffic features used in correlation, stipulating that flow correlation can be performed directly on compressed traffic features instead of on raw traffic features. 
    Lopes et al.~\cite{lopes2024flow} correlate flows based on the similarity of feature vectors (akin to traffic aggregation matrices~\cite{rf}) whose cells contain the number of packets observed within a small time frame. 
    
    In Nasr et al.~\cite{nasr2017compressive} and Lopes et al.~\cite{lopes2024flow}, however, flows' succinct representations are only generated \textit{after} the initial collection of per-packet information. Still, these compact structures may reveal useful for correlation efforts in the scope of stepping-stone detection, should one be able to compute these representations on-the-fly, eschewing the need to store per-packet data. Inspired by these works, we conjecture that these techniques can help reduce memory use at traffic collection nodes and correlate flows using limited flow data (\S\ref{subsec:compact}).

\subsection{P4 Switches as a Platform for Traffic Analysis}
\label{subsec:rw-switches}

This section discusses how programmable switches accelerate traffic analysis in high-speed networks and describes how they have been used for realizing ML-enabled cybersecurity workloads---including network-wide data correlations.

\mypara{Primer on P4-programmable switches.} P4 switches cleanly separate the responsibilities of the network's data and control planes. The data plane is optimized for line-rate packet forwarding and allows for programmable, per-packet operations that enable feature extraction without compromising throughput. In turn, the control plane manages rule installation and updates, supporting adaptive responses to changing traffic patterns.
P4 switches, such as the Intel Tofino or AMD Pensando devices, move packets through a multi-stage pipeline before forwarding them. Ingress and egress pipelines employ \textit{match-action units} to handle packet forwarding and programmable logic. After incoming packets are parsed, their headers and metadata can \textit{match} a given table, whose entry will map to an \textit{action} unit. Actions can alter packet header fields and modify stateful memory (e.g., increment register counters). Although matching tables and other P4 objects are instantiated inside match-action units, they are populated by the control plane at and throughout run-time.

Despite their benefits, P4 switches bring limitations that restrict programmability, including the lack of dynamic data structures, no support for floating-point arithmetic, limited memory capacity ($\sim$256MBs SRAM), and tight computational constraints that allow only simple operations per pipeline stage. 
These constraints pose implementation challenges to P4 programs that require complex flow feature processing and storage, which are typically implemented via clever ``hacks'' and workarounds~\cite{netwarden,barradas2021flowlens,yan_brain-switch_2024}. 
In our proposed design (\S\ref{sec:design}), we leverage similar approaches to make an efficient use of the limited memory and computation primitives in P4 switches to compute and store flow features.

\mypara{Traffic analysis on P4 switches.} P4 switches have been increasingly employed for traffic analysis tasks~\cite{defenseSurvey}, including traffic classification~\cite{barradas2021flowlens}, covert channel detection~\cite{netwarden}, and DDoS mitigation~\cite{patronum}. 
Seminal systems in this space focused on extracting fine-grained traffic features within the data plane, and then offloading them to the switches' control plane to support a range of security-focused tasks. For enhancing traffic analysis capabilities, researchers designed efficient data structures and scalable storage management mechanisms for handling many concurrent flows~\cite{superFe}, as well as making significant strides for running classifiers in the data plane~\cite{yan_brain-switch_2024}.
\textcolor{black}{Despite these strides, flow correlation requires maintaining temporal relationships across multiple flows and associating them with their upstream sources, requiring feature aggregation mechanisms that are not addressed in prior work.}

To the best of our knowledge, the use of P4 switches has not yet been applied to the problem of flow correlation. The closest work to our setting is DELTA~\cite{kirci_mass_2022}, a system where P4-programmable switches are configured to independently identify the establishment of VoIP calls between peers across the network, and then orchestrated to exchange call information to identify the users engaged in communication. However, DELTA does not extract packet or flow-based features that would apply for generic flow correlation tasks. 
This gap presents an opportunity to study whether P4 switches can act as the backbone of scalable and decentralized infrastructure for attack attribution in high-speed networks (\S\ref{sec:design}).

\section{The Attack Attribution Problem}\label{ps}
\label{sec:attribution}

We model a networked environment such as the Internet, consisting of a set of interconnected networks \(\mathcal{N}=\{N_1, N_2, \ldots, N_n\}\). Each network is managed by an operator, e.g., an Internet Service Provider (ISP), a cloud service provider (CSP), or a university campus/enterprise network administrator. These networks may deploy intrusion detection systems (IDSes) in firewalls, servers, or end-hosts, and share telemetry or flow-level metadata to support collaborative cybersecurity operations. We name these \textit{cooperating networks}.

When an IDS signals an attack on a given network \(N_i\), we assume that the objective of the network operator is not merely to identify and respond to the event (e.g., by dropping the offending flow and/or temporarily preventing further communication from a given proxy's IP), but to carry out \textit{attack attribution}, i.e., to uncover the actual IP address $a \in \mathcal{N}_\mathsf{IP}$ launching the the attack.

To perform attack attribution, a network operator extracts traffic features from the malicious flow---denoted \( f_m^{i} \)---as observed in the attacked network \( N_i \). These features are then correlated with those of other flows \( f_k \in \mathcal{F} \) originated within cooperating networks in \( \mathcal{N} \setminus \{N_i\}\).  Here, $f_k^{j} \in \mathcal{F}(N_j, a)$ denotes a flow originated within network $N_j$, and whose source IP is $a$.
Let \( \rho(f_m^{i}, f_k^{j}) \) denote the correlation between the features of \( f_m^{i} \) and \( f_k^{j} \). Correlation may reflect similar traffic timing and volume characteristics, indicating that \( f_k^{j} \) and \( f_m^{i} \) may belong to the same communication path or originate from a common source, as observed from different vantage points.
Hence, \textit{attack attribution} can be formalized as identifying the probable attacker's IP address \( \hat{\mathcal{A}} \), as follows: 

\vspace{-0.3cm}
{\small
\[
\hat{\mathcal{A}} = \arg\max_{a \in \mathcal{N}_\mathsf{IP}} \sum_{N_j \in \mathcal{N} \setminus \{N_i\}} \sum_{f_k^{j} \in \mathcal{F}(N_j, a)} \rho(f_m^{i}, f_k^{j}) \quad \]
} 

\noindent $\text{subject to } \rho(f_m^{i}, f_k^{j}) \geq \eta$, where $\eta$ is a  similarity threshold.

\begin{figure}[!t] \centering \includegraphics[width=\linewidth]{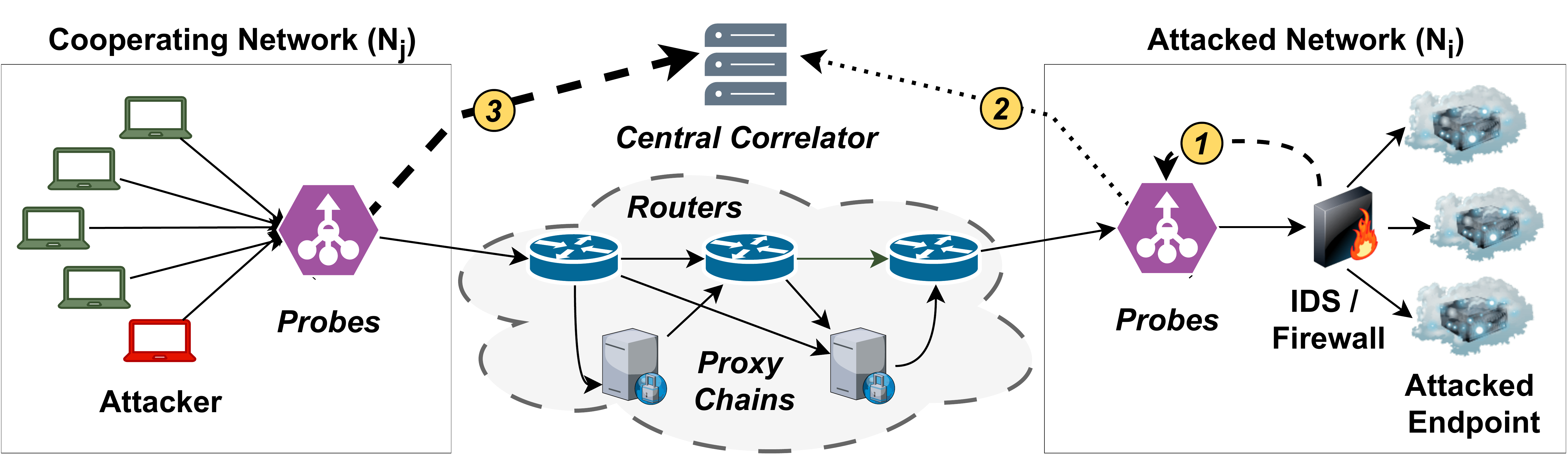} 
\vspace{-0.5cm}
\caption{Centralized correlation design for attack attribution.} \label{fig:central-corr} 
\vspace{-0.5cm}
\end{figure}

\mypara{Scalability challenges of centralized attack attribution.} Figure~\ref{fig:central-corr} shows a centralized flow correlation architecture for attack attribution. The IDS in the attacked network instructs its probe (step~\encircle{1}) to send the attacking flow's feature vector to a central correlator (step~\encircle{2}), while cooperating networks forward feature vectors for all of their outgoing flows (step~\encircle{3}). The correlator then computes similarity scores between the attacking flow and those from cooperating networks.
This design incurs a computation and communication 
in the order of \( |f_k| \) -- the total number of outgoing flows observed by the cooperating networks. Thus, centralized architectures create significant bottlenecks, requiring powerful correlator nodes and high-capacity links. In the next section, we present \sys, a distributed framework that addresses these limitations.

\section{\sys}
\label{sec:design}

This section introduces \sys, a distributed correlation framework aimed at enabling a coalition of operators managing a set of cooperating networks to identify the network location of an attack's perpetrator. \sys leverages P4 switches to simultaneously act as \textit{probe nodes} (responsible for collecting flows' traffic features) and \textit{correlator nodes} (i.e., responsible for computing correlation scores), thus efficiently distributing correlation workloads across the network.


\subsection{Architecture Overview}
\label{subsec:overview}

\begin{figure}[!t] \centering \includegraphics[width=1\linewidth]{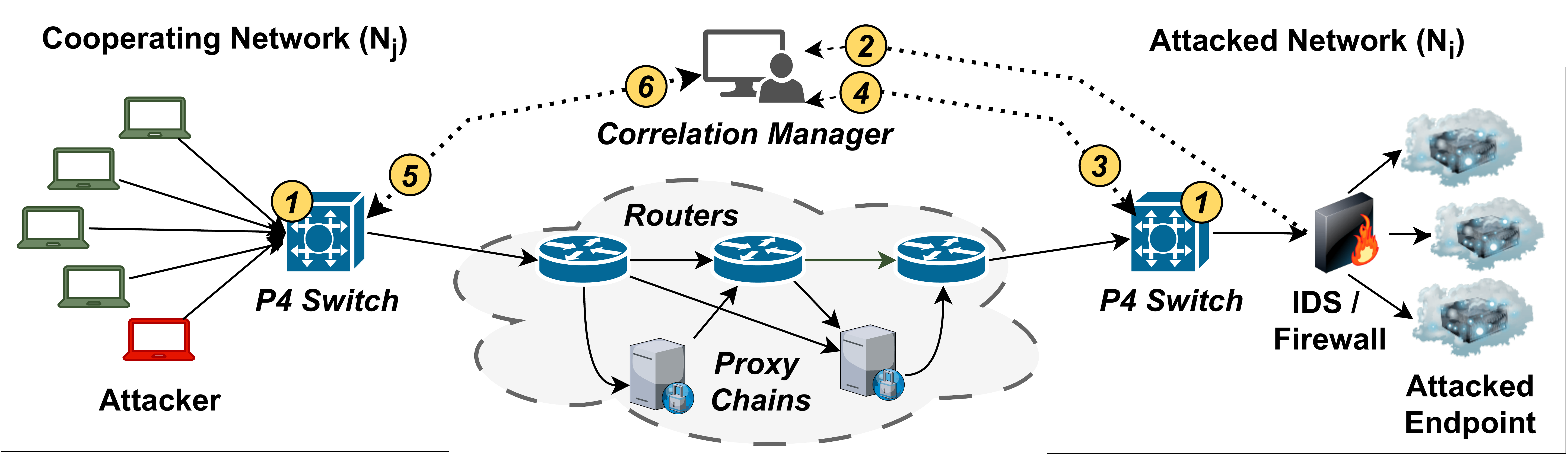} 
\vspace{-0.5cm}
\caption{\sys's decentralized correlation architecture.} 
\label{fig:system_arch} 
\vspace{-0.5cm}
\end{figure}

Figure~\ref{fig:system_arch} illustrates the overall architecture of \sys. The framework is comprised of three key components that operate in tandem. We detail them below.

\mypara{Programmable switches.} P4 switches represent a central component of \sys, serving a dual role as feature collectors (\textit{probe nodes}) and correlation engines (\textit{correlator nodes}). 
Importantly, these switches may already be deployed at participating networks to function as border routers and perform packet forwarding, making them a readily available platform for in-network processing. \sys leverages this existing infrastructure to extract flow-level features in a per-packet fashion, enabling an efficient feature aggregation directly within the data plane (\S\ref{subsec:rw-switches}). 
Once instructed to initiate correlation, the P4 switches retrieve the set of feature vectors associated with each flow of interest from the data plane and perform the correlation operations internally, on the switch's CPU  (control plane). This ensures that feature extraction and correlation remain tightly coupled and execute within the switch itself, requiring no additional components.

\mypara{Intrusion detection systems.} We assume that robust IDSes---such as firewalls, security appliances, or ML-based detectors on middleboxes---are already deployed by a given network's operator. These systems are configured to improve detection against attacks targeting specific services hosted within the network, enhancing their effectiveness. They continuously monitor traffic and trigger alerts upon detecting malicious activity. Each alert sets off a communication with the correlation manager, which then requests the offending flow’s features from the network’s front-facing P4 switch to bootstrap the distributed correlation process.

\mypara{Correlation manager.}
The correlation manager acts as a lightweight logical entity that orchestrates the distributed correlation workload. We envision that this component may be operated independently by a neutral third party, such as a trusted consortium, an inter-organizational security alliance, or a national cybersecurity center~\cite{seldom} \textcolor{black}{for the purpose of attack attribution, e.g., in the context of international law-enforcement collaborations~\cite{untangling}.} This component's role is limited to coordinating the correlation process: it collects metadata and feature vectors of attacking flows from the probe node within an attacked network, and distributes the offending flow's feature vector to the correlator nodes run by cooperating networks. The actual correlation is performed at those distributed nodes, deliberately avoiding centralized resource-intensive computations. 
Once the distributed correlation is complete, results are reported back to the correlation manager.

\textcolor{black}{From a privacy perspective, \sys minimizes data disclosure by revealing a potential attack source IP to the correlation manager only when the correlation score between the malicious flow and a  flow observed at a cooperating network exceeds a predefined threshold. During correlation, only flow sketches are exchanged, preventing the leakage of personally identifiable information that could be reconstructed from detailed flow information. In addition, \sys is compatible with privacy-preserving flow correlation mechanisms, e.g., those based on multi-party computation primitives~\cite{macedo2022privacy}.}

\subsection{Operational Workflow}

We now outline the operational workflow of \sys, relying on
Figure~\ref{fig:system_arch} to illustrate each of these steps.

\mypara{Flow features' extraction.} In steady-state, all probe nodes within a cooperating network will produce a compact representation of the features for each flow concurrently crossing the switch at a given point in time (step \encircle{1}). Given programmable switches' memory limitations, this compact representation, which we refer to as a \textit{feature vector}, is ephemeral,  potentially being replaced in an LRU fashion as flows are terminated~\cite{barradas2021flowlens}. Feature vectors will be used for correlating flows as part of the attack attribution process once an attacking flow is detected within a cooperating network under attack. 
As we describe later on, \sys is compatible with multiple compact representations of a flow's feature vector (\S\ref{subsec:compact}), enabling us to explore different memory/correlation accuracy trade-offs.

\mypara{Attack detection.} The IDS monitors network traffic to identify ongoing attacks. Once an attack is detected, the IDS extracts the 5-tuple information $\langle$\textit{Src. IP, Dst. IP, Src. Port, Dst. Port, Proto}$\rangle$ of the malicious flow, and communicates this data to the correlation manager (step \encircle{2}).
    
\mypara{Request of attacking flow's features.} Upon receiving an alert, the correlation manager requests the feature vector tied to the attacking 5-tuple from the probe node (i.e., P4 switch) deployed on the attacked network (step \encircle{3}).

\mypara{Propagation of the attacking flows' features.} As requested by the correlation manager, the probe node within the attacked network sends the correlation manager a feature vector that characterizes the attacking flow (step \encircle{4}).

\mypara{Correlation directive.} Upon receiving the feature vector of the attacking flow, the correlation manager forwards it to the P4 switches (which will now act as correlator nodes) that front-face each cooperating network. In addition, the correlation manager distributes the attacking flow's details (i.e., the flow's start times and communication volume) to the same switches, enabling them to preemptively identify which (outgoing) recorded flows have similar start/end times as the offending flow, and reason about heuristic optimizations for the local correlation procedure (\S\ref{subsec:optimizations}). Finally, the correlation manager instructs these switches to initiate their local correlation process (\S\ref{subsec:compact}) for the attacking flow (step \encircle{5}).

\begin{figure}[!t] \centering \includegraphics[width=\linewidth]{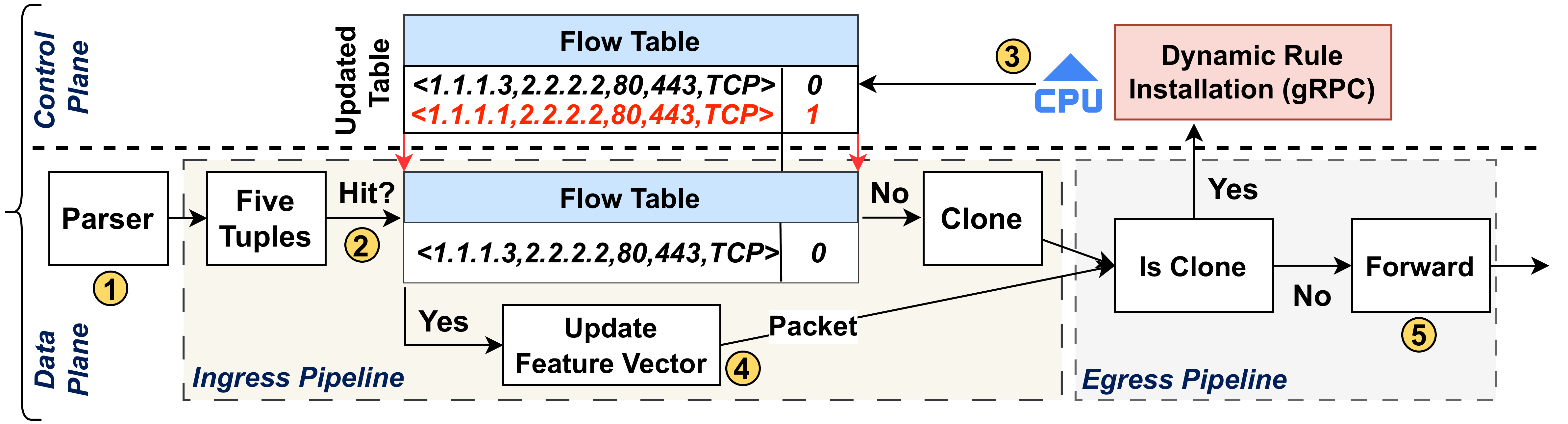} 
\vspace{-0.4cm}
\caption{\sys's dynamic flow identification mechanism.} \label{fig:row} 
\vspace{-0.5cm}
\end{figure}

\mypara{Flow correlation and reporting.} Flows with correlation scores matching and/or exceeding a set threshold are flagged as potentially correlated, and any associated 5-tuples are sent back to the correlation manager for concluding the attack attribution process (step \encircle{6}). 
\textcolor{black}{The manager performs the final attack attribution by aggregating reports from multiple P4 switches (scattered across several networks and sub-networks), identifying candidate attack sources. It could also partially or even fully reconstruct the path of the malicious traffic, even if intermediate cooperating networks did not detect the attack.}

\mypara{A note on insider attacks.} \textcolor{black}{Insider attacks represent a simplified case of \sys's workflow. Correlation could be performed within the P4 switching infrastructure of the ISP, assisted by an in-house correlation manager, without the need to share sketches externally. However, correlation accuracy improves with broader visibility, as prior traffic correlation work has shown~\cite{lopes2024flow}; restricting correlation to a single network risks missing the true correlated flow if it transits elsewhere. In such cases, weaker local matches may be mistaken for correlations, reducing true positives and increasing false positives.}

\subsection{Flow Identification and Tracking}
\label{subsec:flow-id}

A key operation underlying the execution of \sys is that of performing an efficient flow identification within P4 switches, so that accurate per-flow feature vectors can be computed and stored within each switch's data plane. Still, previous prototypes for ML-based flow analysis schemes for P4 switches side-step the problem of unique flow identification, assuming that some process is in place to perform it (e.g.,~\cite{barradas2021flowlens}), or forcing the initial packet of a new connection to move through the control plane for further processing and mapping (e.g.,~\cite{netwarden}), thus causing delays upon connection establishment in high-speed networks with low latency requirements. Thus, while apparently straightforward, producing a 1-to-1 match between new flows and a memory region that can accommodate for a flow's compact feature representation is non-trivial should one wish to avoid delays in packet processing and/or feature corruption, e.g., caused by hash-based indexing methods that may lead to collisions~\cite{p4ddle}.

\mypara{\sys's flow identification pipeline.} We now detail how \sys tackles flow identification, relying on Figure~\ref{fig:row}.

\begin{compactitem}
\item \textit{Packet parsing.} The first step on the packet processing pipeline involves a parsing operation that extracts the packet's 5-tuple (step \encircle{1}).

\item \textit{Flow table lookup.} We introduce a dedicated \textit{flow table} that stores reference indices pointing to row entries in another data plane \textit{feature table}. The latter logically organizes registers' memory in rows, where each row stores a given flow's feature vector. After parsing, a packet's 5-tuple is checked against the flow table (step~\encircle{2}). We must install a new rule for each newly observed flow.

\item \textit{Packet cloning and rule installation.} To install a rule for a new flow, the first packet is cloned: the original is forwarded without delay, while the clone is sent to the control plane for rule creation. The rule is installed in the data plane's flow table through a remote procedure call (step~\encircle{3}), and will point to a cell in the feature table for keeping track of packets pertaining to the new flow.
Note that a flow's initial packet(s) triggering a rule installation will not be included in a flow's feature vector, since the index to write on the data plane is not yet available. In this case, multiple packets from the same flow may be temporarily queued before the rule is installed;  
accordingly, only the first packet triggers rule installation, while subsequent packets are ignored.

\item \textit{Feature vector updates and packet forwarding.} After rule installation, the feature vector is updated as subsequent packets from the flow pass through the switch (step~\encircle{4}). Once a packet is fully processed, it is forwarded to the appropriate port as dictated by the switch's IPv4 forwarding table (step~\encircle{5}).
    
\end{compactitem}

\subsection{Compact Flow Features' Representation and Correlation}
\label{subsec:compact}

Traffic correlation techniques target a set of features that are commonly derived from per-packet information and which remain invariant over time. Examples include packet timestamps, sizes, and overall communication volume. Storing a flow's per-packet information on a programmable switch's data plane, however, would occupy a significant memory footprint, impacting the total amount of concurrent flows that could be correlated by \sys at any given time. 

To minimize the amount of data that must be kept by a switch for each active flow, we explore existing methods of generating compact feature vectors. These have been found to be applicable to traffic correlation workloads as well as other ML-based security tasks focused on flow analysis.

\mypara{Traffic aggregation matrix.} We first adopt a methodology (for generating feature vectors) which builds a traffic aggregation matrix (TAM)~\cite{rf}. This matrix records metadata about packets transmitted per flow across multiple bins of $t$ seconds each, for a maximum of $T$ seconds, thereby storing a flow-level feature vector in a fixed-size data structure, consuming significantly less memory when compared to storing individual packet features. Like Lopes et al.~\cite{lopes2024flow}, we generate a single-row TAM per flow, where each TAM bin tracks the number of packets transmitted by a flow within that bin's time interval.

While storing complete TAM feature vectors in a switch's data plane might be feasible, their memory footprint may compromise the concurrent storage of many flows simultaneously crossing the switch (see TAMs' trade-offs in \S\ref{sec:results}).

\begin{table*}[t]
  \centering
  \small
  \caption{Storage requirements in terms of \(f\) (number of flows), \(n\) (TAM length), and \(m\) (sketch length). Integers are 32 bits.}
  \label{tab:storage2}
  \resizebox{0.82\linewidth}{!}{%
  \begin{tabular}{L{3.8cm} | L{2.7cm} L{2.7cm} | L{3cm} L{4.5cm}}
    \toprule
    \textbf{Method / Storage} & \textbf{Proj. Matrix} & \textbf{Flows} & \textbf{Total (bits)} & \textbf{Total (bits) w/ heur. (\S\ref{subsec:optimizations})} \\
    \midrule
    Nasr et al. (integer sketch) & 
    \(n \times m\) (integers) & \(f \times m\) (integers) &
    \(32\,(n \times m + f \times m)\) &
    \(32\,(n \times m + f \times m + f) + 48 \times f\) \\

    Coskun et al. (integer sketch) & 
    \(n \times m\) (integers) & \(f \times m\) (integers) &
    \(32\,(n \times m + f \times m)\) &
    \(32\,(n \times m + f \times m + f) + 48 \times f\) \\

    Coskun et al. (binary sketch) & 
    \(n \times m\) (integers) & \(f \times m\) (bits) &
    \(32\,(n \times m) + f \times m\) &
    \(32\,(n \times m + f) + f \times m + 48 \times f\)  \\ \hline

    TAM feature vector & 
    -- & \(f \times n\) (integers) &
    \(32\,(f \times n)\) &
    \(32\,(f \times n + f) + 48 \times f\) \\
    \bottomrule
  \end{tabular}
  }\vspace{-0.5cm}
\end{table*}

\mypara{Flow sketching.} To further compress TAM feature vectors while  retaining flows' characteristics, we use sketching techniques based on vector projection methods~\cite{coskun2009online,nasr2017compressive}. Briefly, let the TAM feature vector for a flow be $\mathbf{f} = [f_1, f_2, \ldots, f_n]$. Sketching algorithms transform $\mathbf{f}$ into a lower-dimensional vector $\mathbf{f}_c = [f_{c1}, f_{c2}, \ldots, f_{cm}]$, where $m \ll n$. 
The parameterization of such sketches enables us to trade-off the usage of switch memory (and thus, the number of concurrent flows that can be measured) with correlation accuracy.

     We integrate these constructs into \sys's data plane processing logic, contrasting the use of two prominent sketching algorithms, proposed in the traffic correlation literature (\S\ref{subsec:rw-correlation}), as the main driver of \sys's attack attribution mechanisms. We describe them below, and detail how flows' sketches can be compared towards realizing flow correlation.

     \myparait{Coskun et al.~\cite{coskun2009online}} propose an online sketching method that first bins packets into discrete time slots (i.e., a packet count-based TAM vector) and then leverages linear transformations to generate a compact integer-array sketch representation of a flow. Sketches are computed on-the-fly without the need for temporarily storing the complete TAM feature vector. 
     As the basis for these transformations, we use a random projection matrix whose entries are independently drawn from a Bernoulli distribution (i.e., each entry is either +1 or -1 with equal probability). This projection preserves the structure of the packet-count vector and produces a sketch that contains only integer values, offering low per-packet overhead and robustness to network perturbations.
     The sketches can be binarized to save space and enable more efficient comparisons.

\myparait{Nasr et al.~\cite{nasr2017compressive}} propose the aggregation of raw traffic features into a feature vector, which is then compressed using a sensing matrix \(\Phi \in \mathbb{R}^{m \times n}\) into a lower-dimensional sketch. \(\Phi\) satisfies the restricted isometry property, allowing Euclidean distances between features to be preserved in the compressed domain.

In our implementation, we conducted two adaptations to Nasr et al.'s~\cite{nasr2017compressive} original approach. First, since this scheme originally compresses the full feature vector, we implement a continuous update of a flow's feature sketch every time a packet is processed, thus replicating the online sketching nature of Coskun et al.~\cite{coskun2009online}. Second, since sensing matrices \(\Phi\) are instantiated as random Gaussian matrices with std. dev. \(\sigma = 1\), these contain floating-point entries which are not supported by P4 switches (\S\ref{subsec:rw-switches}). To address this issue, we scale the matrix \(\Phi\) by a constant factor---\(10\,000\), in our implementation---to convert its entries to integer values without losing significant precision. Sketching is then performed on the P4 switch using this scaled matrix, enabling integer-only arithmetic.

Table~\ref{tab:storage2} depicts the storage requirements for holding a TAM for a single flow, when contrasted to the storage required to hold the sketches we consider~\cite{coskun2009online,nasr2017compressive}.

\mypara{Correlation.} 
The final step in \sys' pipeline involves correlating feature vectors to identify whether two flows collected at different vantage points originate from the same source. In \sys, correlation is based on computing a statistical distance or similarity between the sketches. 
Different sketching methods use disparate metrics for enacting said comparisons. Coskun et al.~\cite{coskun2009online} use the Hamming distance to compare sketches while Nasr et al.~\cite{nasr2017compressive} employ cosine similarity. Our implementation relies on the same metrics.

\subsection{Heuristic Optimizations for Attack Attribution} 
\label{subsec:optimizations}

While the above correlation methods provide a starting point for \sys's attack attribution, probe nodes in cooperating networks observe a large volume of unrelated flows. These unrelated flows increase correlation complexity and the risk of false positives, as noted in prior work~\cite{coskun2009online,nasr2017compressive}. To address this, we adopt two optimizations~\cite{blum2004detection} that reduce the flow search space at each correlator node.

\mypara{Creation time heuristic.} Since correlation targets flows that occurred within a small interval relative to the attacking flow, we exclude flows whose start times fall outside a temporal window defined w.r.t. the start time of the attacking flow (as tracked at the attacked network). Thus, we only consider flows with initial timestamps within \(T_{\min}\) and \(T_{\max}\), offset from the attacking flow’s start time. 

\mypara{Packet count heuristic.} Flows with packet counts akin to that of the malicious flow are more likely to be true matches. By bounding the acceptable packet count range using thresholds \(P_{\min}\) and \(P_{\max}\), derived from the target malicious flow, we restrict correlation to flows with comparable traffic volumes.

We apply the heuristics one after the other.
This two-step strategy reduces correlation complexity from the baseline \( O(|\mathcal{F}| \times C) \)—where \( |\mathcal{F}| \) is the total number of outgoing flows observed in $\mathcal{N} \setminus \{N_i\}$ (see \S\ref{sec:attribution}) and \( C \) the cost of a single comparison between two flows' feature vectors—to \( O\left(\log(|\mathcal{F}|) + |f_t| + |f_{t+p}| \times C\right) \). Here, \( f_t \) and \( f_{p+t} \) represent the reduced flow sets after the cumulative timestamp and packet count filtering, respectively. Since flows are pre-sorted by timestamp, identifying \( f_t \) requires only \( O(\log(|\mathcal{F}|)) \) via binary search. Filtering by packet count is linear in \( f_t \), yielding \( f_{t+p} \) in \( O(|f_t|) \). We then perform flows' feature vector comparisons only on this set, which incurs a cost of \( O(|f_{t+p}| \times C) \).

To implement the heuristics, the data plane of each switch maintains two separate tables with a number of rows equal to the number of flows. Each entry stores auxiliary metadata: 48 bits for a flow's creation timestamp and 32 bits for that flow's total packet count. This results in a storage overhead of \(32 \times f + 48 \times f\), where \(f\) denotes the number of flows observed by a switch (see Table~\ref{tab:storage2}).

\subsection{Implementation}

We built a prototype of \sys using \texttt{bmv2}, the reference P4 software switch. The data plane logic, including flow identification and sketching operations (for either sketch), was implemented in $\sim$500 lines of P4$_{16}$ code. In turn, \sys's control plane logic was written in $\sim$300 lines of Python code. This includes the installation of tables and rules supporting flow identification and sketching operations in the data plane, as well as fetching flows' feature vectors via reads to data plane registers for enabling the correlation backbone.

\section{Evaluation}
\label{sec:eval}

\subsection{Evaluation Goals and Metrics}\label{goals} 

We assess \sys's practicality along three facets:

\mypara{Effectiveness.} We evaluate \sys's attack attribution capability by measuring its correlation accuracy on malicious flows, using metrics aimed to capture the trade-off between successful correlations and incorrectly matched flows~\cite{nasr2018deepcorr}.

The \textit{true positive rate} (TPR) measures the fraction of attacking flows that are correctly correlated by the system. Let \( f_m^{s} \) denote the number of malicious flows originated within a network with border switch \( s\), and \( TP^{s} \) the number of those that are correctly matched to the malicious flows detected within a cooperating network under attack. Across all switches \( \mathcal{S} \), \( \text{TPR} = \sum_{s \in \mathcal{S}} TP^{s} \big/ \sum_s f_m^{s} \).

The \textit{false positive rate} (FPR) captures incorrect correlations. Let \( f^{s} \) denote the total number of flows originated within a network with border switch \( s \), and \( FP^{s} \) the number of such flows that are incorrectly matched to malicious flows $f_m$ detected within an attacked network. Then \( \text{FPR} = \sum_{s \in \mathcal{S}} FP^{s} \big/ \sum_s f_m f^{s} \), where the denominator reflects all potential false-positive pairs across all switches.

\mypara{Efficiency.} We quantify the computational cost linked to flow correlation via the number of pairwise comparisons between detected malicious flows and the outgoing flows observed by cooperating networks. The computational effort across all switches \( \mathcal{S} \) is expressed as the sum of the pairwise comparisons for each switch: \( \sum_{s \in \mathcal{S}} f_m \times f^{s} \), where \( f_m \) represents the number of malicious flows detected and \( f^{s} \) represents the number of outgoing flows observed by switch \( s \).

\mypara{Scalability.} We evaluate \sys' scalability by analyzing two key factors: a) the number of flows that can be concurrently stored and processed, and; b) the communication overhead required during attack attribution. Let \( f^{s} \) denote the number of outgoing flows observed by a cooperating network's switch, and let \( C^{s} \) represent the total communication cost (in bits) for transmitting these flows' features'. If each flow is represented by a feature vector of size \( m \) bits, then transmitting all flows \( f^{s} \) incurs a communication cost of \( C^{s} = f^{s} \times m \).

\subsection{Evaluation Methodology}

We now describe the datasets used in our evaluation, and the configuration of \sys's data structures and heuristics.

\mypara{Datasets used for attack attribution.} We use six labelled network traffic datasets, released by Fu et al.~\cite{fu_detecting_2023}, as a target of \sys' attack attribution capabilities. These datasets were compiled from a combination of Fu et al.'s own experimental data and traffic traces from the WIDE MAWI project in Tokyo, Japan. For exercising 
\sys's generalizability, we selected each dataset from six different categories of attacks collected by Fu et al. (see Table~\ref{tab:selected_datasets} for a description). Each dataset contains a different type of network attack along with background benign traffic. Fu et al.'s data details per-packet five-tuples in \texttt{.csv} files, along with packets' timestamps and labels (benign/malicious), allowing us to carve out individual flows identified by these 5-tuples. Each dataset accounts for more than 100k flows, and the ratio of benign to malicious traffic is at least 39:1 (Dridex) and at most 1312:1 (Oracle). 

For simplicity, we assume that the network IDS deployed within each \sys-enabled network acts as an oracle that can perfectly distinguish between benign and malicious flows. While this assumption is already aligned with the capabilities of state-of-the-art IDSes for the datasets we considered~\cite{fu_detecting_2023}, we recall that our goal is not to perform accurate malware classification, but rather to act on IDSes' alerts (\S\ref{sec:attribution}). \textcolor{black}{In practice, false positives would increase the number of correlation tasks to be performed, while false negatives would prevent the attribution of some attacks, since those malicious flows would never be sent to the correlation manager.}


\mypara{Simulating vantage points and network conditions.} Since the above data\-sets were collected at a single network vantage point and do not include raw packet traces that can be transparently replayed across some network topology (real or emulated) by special-purpose software such as \texttt{tcpreplay}, they cannot be directly used for correlation experiments across different networks, as required by \sys. 

To tackle this issue, akin to~\cite{coskun2009online}, we simulate the acquisition of two separate observations for each flow at different vantage points within \sys-enabled networks: a) at the border router of a cooperating network where hosts originate benign/malicious traffic, and; b) at the border router of a network which is targeted by some attack. We also assume that all flows in each dataset originate from a cooperating network and traverse (or target devices within) the attacked network.

To facilitate this setup, we implemented a simulator that models WAN traffic relayed via a proxy node. 
We used the simulator to augment the traffic traces of Fu et al.~\cite{fu_detecting_2023}, reproducing the observation of flows across two vantage points to mimic distributed monitoring. The resulting traces, which we use throughout our evaluation, capture traffic across the WAN where packets incur an average latency increase of $\sim$200ms between any two vantage points; \textcolor{black}{this falls within the range of end-to-end latencies between geo-distributed client-proxy-server nodes in stepping-stone scenarios that consider wide-area paths~\cite{webb2016finding}.} The simulator also supports the injection of packet losses for us to assess the robustness of flow correlation under network perturbations.

\mypara{Parameterization of \sys's sketches.}
Each dataset from Fu et al.~\cite{fu_detecting_2023} spans 45--65s of traffic. To explore the impact of temporal granularity in flow feature collection, we generated TAM time bins (\( t \)) of 0.1s, 0.5s, and 1s. 
We performed preliminary experiments using different sketch lengths (\( m = 5, 10, 15 \)), keeping \( m = 10 \) as a baseline. A sketch length of 5 slightly improved TPR by up to +1.52\% but at the cost of a substantial increase in FPR, reaching +114.84\%. Conversely, using \( m = 15 \) provided no consistent TPR gains and resulted in mixed FPR outcomes (ranging from –20.4\% to +92.65\%), along with added storage overhead. Overall, for all datasets we considered, \( m = 10 \) strikes a favourable balance between accuracy and efficiency (\S\ref{sec:results}). Still, this parameter may need to be tuned for flows with different traits~\cite{coskun2009online}.

We follow the original methodology of each sketch to compute correlation scores.  For Coskun et al.'s sketch, we use Hamming distance and consider a match to be a true positive only when the distance between sketches is 0. For Nasr et al.'s sketch, we use cosine similarity, requiring a score of 1 for an exact match. We evaluate TAMs with both Hamming distance and cosine similarity, applying the same thresholds to define true positives. We adopt these thresholds to reflect high-confidence matches in attack attribution, where false associations can be harmful to benign users.
Indeed, we experimented with relaxed correlation thresholds across various time bins $t$, but found that these looser criteria offered only marginal improvements in TPR while significantly increasing FPR. For instance, for Nasr et al.'s sketches, a cosine similarity threshold of 0.9 yielded modest increases in TPR (up to 2.54\%) but significantly higher FPR (up to 373.04\%).

\begin{table}[t]
  \centering
  \caption{Network traffic datasets used in our experiments.}
  \label{tab:selected_datasets}
  \resizebox{\linewidth}{!}{%
  \begin{tabular}{L{1.6cm} L{1.5cm} L{3.5cm} r @{\hspace{1em}} r r}
    \toprule
    \textbf{Dataset} & \textbf{Category} & \textbf{Description} & \multicolumn{2}{c}{\textbf{Flows}} & \textbf{Span (s)}\\
    \cmidrule(lr){4-5}
    & & & \textbf{Benign} & \textbf{Malicious} & \\
    \midrule
    Snojan       & Botware     & PPI malware downloading.           & 206\,723 & 1\,607  & 45.64 \\
    Dridex       & Ransomware  & Victim locations uploading.        & 125\,424 & 3\,202  & 54.75 \\
    Adload       & Adware      & Resources for PPI adware.   & 125\,417 & 602    & 54.80 \\
    Oracle       & Web         & TLS padding Oracle.                & 294\,110 & 224    & 64.14 \\
    Penetho      & Spyware     & Wifi cracking APK spyware.         & 293\,808 & 1\,006  & 55.64 \\
    Bitcoinminer & Miner       & Abnormal encrypted channels.       & 125\,418 & 202    & 61.01 \\
    \bottomrule
  \end{tabular}
  }\vspace{-0.5cm}
\end{table}

\mypara{Configuration of \sys's heuristics.} Network topology and flows' characteristics can influence \sys's heuristics' configurations~\cite{blum2004detection}. However, since the datasets we consider (see Table~\ref{tab:selected_datasets}) share similar traits on flow durations, we configure our heuristics to be consistent across all datasets.
For the timing-based and packet count heuristics, we empirically found that a window of $\pm 2.5$ seconds and a threshold of $\pm 5\%$ traffic volume filters out irrelevant candidate flows while retaining most valid matches. As described in \S\ref{sec:results}, applying these heuristics to \texttt{\small bitcoinminer} reduces the number of comparisons per attacking flow by 3 orders of magnitude, without substantially sacrificing correlation accuracy.

\subsection{Evaluation Results}
\label{sec:results}

We now describe the main results obtained during our evaluation of \sys. We centre our exposition on the \texttt{\small bitcoinminer} dataset since it: a) has a representative benign-to-malicious flow ratio of 625:1, close to the median of all datasets; b) exhibits the second-largest temporal span, and; c) has a relatively smaller number of flows, facilitating faster experimentation. Results for other datasets exhibit similar trends, and we defer their discussion to Appendix~\ref{app}.

\begin{figure}[!t]
\centering    \includegraphics[width=0.9\linewidth]{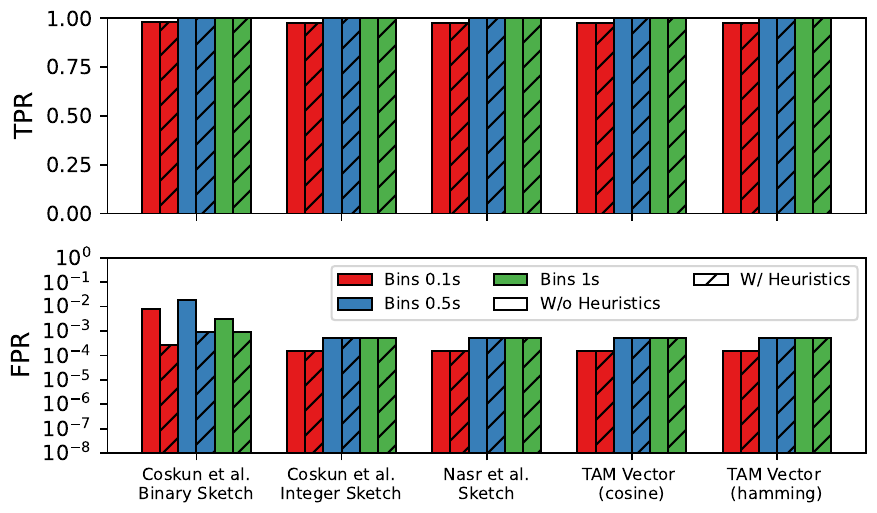}
    \vspace{-0.2cm}
    \caption{\texttt{\small bitcoinminer} correlation scores for different sketch and TAM configurations (for unperturbed network conditions).}
\label{fig:bitcoinminer_comparison}
    \vspace{-0.4cm}
\end{figure}

\begin{figure}[!t] \centering \includegraphics[width=0.9\linewidth]{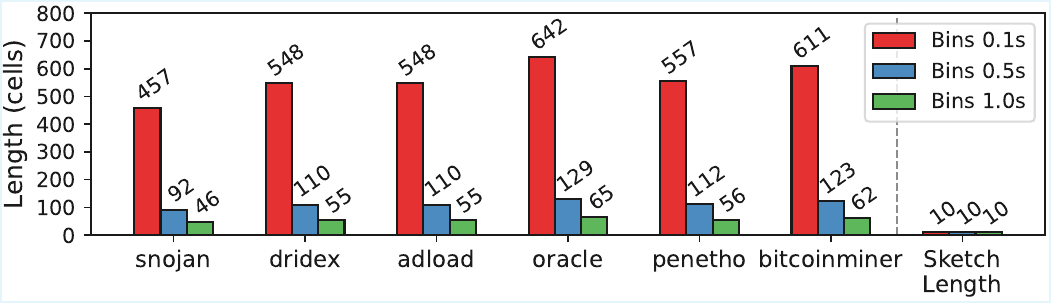} 
\vspace{-0.2cm}
\caption{Length of TAM feature vectors vs. sketches' length.} 
\label{fig:sketch-len} 
\vspace{-0.5cm}
\end{figure}

\mypara{Compact sketches obtain correlation accuracy equivalent to TAMs.}
Figure~\ref{fig:bitcoinminer_comparison} shows the TPR and FPR achieved by different parameterizations of the TAM and sketching approaches considered in our work, when correlating flows pertaining to the \texttt{\small bitcoinminer} dataset. The figure suggests that sketches attain a comparable correlation performance vs. TAMs, making them highly attractive due to their smaller memory overhead. Focusing on the results obtained without the use of heuristics (solid bars), the integer sketching method from Coskun et al.~\cite{coskun2009online} achieves equivalent accuracy (TPR: $0.9917$, FPR: $3.98 \times 10^{-4}$) to TAM (in 0.1 seconds time bin and Hamming distance) while offering significant space savings---indeed, Figure~\ref{fig:sketch-len} illustrates that TAM's memory footprint can be up to $60\times$ larger than that of sketches under finer-grained binnings (e.g., $t=0.1$) across all datasets.

\begin{figure}[!t]
    \centering
 \includegraphics[width=0.9\linewidth]{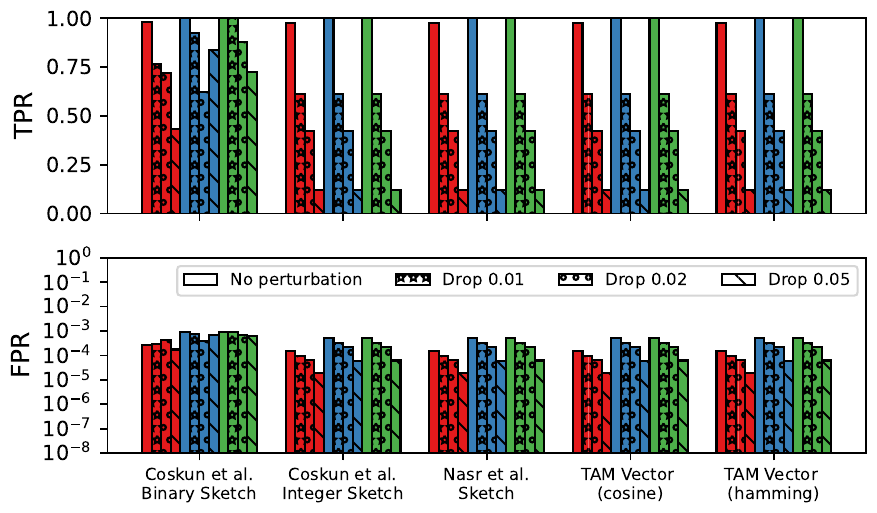} 
    \vspace{-0.2cm}
    \caption{\texttt{\small bitcoinminer} correlation scores for different sketch and TAM configurations (for different packet loss ratios).}
    \label{fig:second_comparison}
    \vspace{-0.4cm}
\end{figure}

\begin{table}[b]
    \vspace{-0.3cm}
    \centering
    \begin{minipage}{0.49\linewidth}
        \centering
\caption{Feature vector comparisons for the heuristics.}
\label{tab:averages_summary}
\resizebox{\linewidth}{!}{%
\begin{tabular}{l @{\hspace{1em}} r @{\hspace{1em}} r @{\hspace{1em}} r @{\hspace{1em}} r}
  \toprule
  \textbf{Dataset} & \textbf{None} & \makecell{\textbf{Creation} \\ \textbf{Time}} & \makecell{\textbf{Packet} \\ \textbf{Count}} & \textbf{Both} \\
  \midrule
  Dridex        & 128{\,}626 & 19{\,}409 & 2{\,}403 & 824 \\
  Adload        & 126{\,}019 & 17{\,}386 & 1{\,}168 & 258 \\
  Snojan        & 208{\,}330 & 20{\,}975 & 2{\,}735 & 693 \\
  Oracle        & 294{\,}334 & 41{\,}580 & 95      & 84  \\
  Bitcoinminer  & 125{\,}620 & 17{\,}288 & 791     & 207 \\
  Penetho       & 294{\,}814 & 28{\,}040 & 1{\,}735 & 454 \\
  \bottomrule
\end{tabular}
}
\end{minipage}
    \hfill
    \begin{minipage}{0.48\linewidth}
    \caption{Total flows that can be stored in a P4 switch.}
\label{tab:maxflows_reduced}
\resizebox{\linewidth}{!}{%
\begin{tabular}{l @{\hspace{1em}} r @{\hspace{1em}} r}
  \toprule
  \textbf{Method/ Storage} & \makecell{\textbf{Per flow} \\ \textbf{(in Bytes)}} & \makecell{\textbf{Stored Flows} \\ \textbf{(in 256 MB)}} \\
  \midrule

  Coskun et al. (bin.) & 1.25 & 
  \(\approx 204.8\times10^6\) \\
  Coskun et al. (int.) & 40 & 
   \(\approx 6.4 \times 10^6\) \\
  Nasr et al. (int.) & 40 & 
  \(\approx 6.4 \times 10^6\) \\\hline
  
  TAM (0.1s bins) & 2864 & 
  \(\approx 1.05 \times 10^5\) \\
  TAM (0.5s bins) & 576 & 
  \(\approx 5.20 \times 10^5\) \\
  TAM (1s bins) & 291 & 
  \(\approx 1.03 \times 10^6\) \\
  \bottomrule
\end{tabular}
}
\end{minipage}
\end{table}

\begin{table*}[t]
  \centering
  \small
  \caption{Communication overhead (in bits) for centralized and distributed correlation, evaluated per sketch (with heuristics) in the \texttt{\small bitcoinminer} dataset. The last column shows the overhead reduction under \sys's distributed setup.}
  \label{tab:bandwidth_summary}
  \resizebox{0.8\linewidth}{!}{%
  \begin{tabular}{@{}l
                  | r @{\hspace{2em}}  r @{\hspace{1em}} r @{\hspace{1em}} r |@{\hspace{1em}}
                  r @{\hspace{1em}} r @{\hspace{1em}} r @{\hspace{1em}} r
                  |r@{}}
    \toprule
    \textbf{Method} &
    \multicolumn{4}{c}{\textbf{Centralized}} &
    \multicolumn{4}{c}{\textbf{Distributed (\sys)}} &
    \textbf{OH Red. (\%)} \\
    \cmidrule(lr){2-5} \cmidrule(lr){6-9}
    &
    \textbf{Sw$_c$ $\rightarrow$ CS} &
    \textbf{Sw$_a$ $\rightarrow$ CS} &
    &
    \textbf{Total} &
    \textbf{Sw$_a$ $\rightarrow$ CM} &
    \textbf{CM $\rightarrow$ Sw$_c$} &
    \textbf{Sw$_c$ $\rightarrow$ CM} &
    \textbf{Total} &
    \\
    \midrule

    Coskun et al. (bin.) &
    1\,193\,390 &
    2\,020 &
    & 1\,195\,410 &
    2\,020 &
    38\,380 &
    736\,896 &
    777\,296 &
    35.0\% \\

    Coskun et al. (int.) &
    38\,188\,480 &
    64\,640 &
    & 38\,253\,120 &
    64\,640 &
    1\,227\,680 &
    736\,896 &
    2\,029\,216 &
    94.7\% \\

    Nasr et al. (int.) &
    38\,188\,480 &
    64\,640 &
    & 38\,253\,120 &
    64\,640 &
    1\,227\,680 &
    736\,896 &
    2\,029\,216 &
    94.7\% \\\hline

    TAM &
    229\,085\,280 &
    387\,840 &
    & 229\,473\,120 &
    387\,840 &
    7\,368\,960 &
    736\,896 &
    8\,493\,696 &
    96.3\% \\
    
    \bottomrule
  \end{tabular}
  }\vspace{-0.5cm}
\end{table*}

To assess if sketches can succeed under perturbed network conditions, we simulated random packet drops at $1\%$--$5\%$ rates. We now present our findings assuming that heuristics were deployed. As shown in Figure~\ref{fig:second_comparison}, both TPR and FPR declined under packet loss (bars with markers), consistent with prior observations~\cite{coskun2009online,nasr2017compressive}. For instance, the integer sketch ($t=$0.1s) maintained a TPR of $0.1188$ and FPR of $4.7 \times 10^{-5}$, equivalent to that of TAM {under the packet drop of 5\%}. The binary sketching variant yielded a significantly higher TPR ($0.6634$) compared to both integer sketches and TAM in the same conditions, albeit with a modest increase in FPR. 

Overall, the above results suggest that, even in noisy environments, sketches preserve the detection characteristics of more resource-intensive TAM configurations and can be relied upon for realizing \sys's correlation backbone. Yet, advanced traffic shaping techniques beyond network perturbations, such as traffic shaping proxies~\cite{nunes2023enhancing}, present an inherent limitation to traffic correlation schemes, making this process significantly harder, if not infeasible~\cite{oh2022deepcoffea,lopes2024flow}.

\mypara{Heuristics reduce complexity and foster improved correlation.} 
Table~\ref{tab:averages_summary} illustrates the impact of heuristics on reducing the flow comparisons performed by \sys. Without heuristics, comparisons range from 125k (\texttt{\small bitcoinminer}) to nearly 295k (\texttt{penetho} and \texttt{oracle}). With heuristics applied, this drops to 207, 454, and 84, respectively, thus reducing the computational effort involved in the correlation workload by at least three orders of magnitude.

Beyond decreasing computational complexity, the heuristics also substantially lower false positives (see Figure~\ref{fig:bitcoinminer_comparison} -- bars with stripes). For instance, Coskun et al.'s binary sketch (based on packet counts tracked with $t=0.1$) experiences a 96\% decrease in FPR---from \( 0.008 \) to \( 0.0003 \)---after heuristic filtering. This stems from eliminating benign or mismatched attacking flows that appear similar in sketch form but that differ significantly in creation time or total traffic volume.

\mypara{Sketches allow for storing more flows concurrently.}
Table~\ref{tab:maxflows_reduced} presents the approximated number of flows that can be stored in a Tofino v1 P4 switch equipped with $\sim$256 MB of SRAM, for the various feature extraction methods under analysis. All sketches are configured with a length of \( m = 10 \). For Coskun et al. and Nasr et al., the sketching process requires storing a projection matrix of size \( n \times m \) (\S\ref{subsec:compact}), which introduces a storage overhead of 24\,400B, 4\,920B, and 2\,480B for TAMs based on 0.1s, 0.5s, and 1s bins ($t$), respectively. 

Though sketches require this fixed overhead, they dramatically improve storage capacity. Coskun et al.'s binary sketch stores up to \( 204.8 \times 10^6 \) flows, compared to just \( 1.05 \times 10^5 \) with TAM at $t=0.1$s granularity---our most memory-intensive setting. Other sketches show similar scalability, reinforcing that sketch-based correlation is well-suited for memory-constrained P4 switches that must handle large flow volumes.

\mypara{Distributed correlation saves bandwidth.} 
Attack attribution spans multiple P4 switches distributed across different cooperating networks. Thus, correlation scales in an ``embarrassingly parallel'' fashion: each switch handles its local traffic and performs correlation independently. We now gauge the communication overheads imposed by \sys, comparing them to those of centralized attack attribution deployments.

Recall that \sys reverses the traditional data-sharing model of centralized systems, which require all probe nodes (\textit{Sw$_c$}) to send full flow feature vectors to a central server (\textit{CS}), resulting in high bandwidth overhead. Instead, \sys transmits only the feature vectors of \emph{attacking flows}--collected at the attacked network's switch (\textit{Sw$_a$})--to a central correlation manager (\textit{CM}), which then relays them to \sys-enabled switches (\textit{Sw$_c$}) for localized correlation.

To gauge the communication overhead of centralized vs. distributed correlation, we simulate a topology with 20 \sys switches: 19 monitoring outgoing flows at cooperating networks (\textit{Sw$_c$}), and one observing incoming flows at an attacked network (\textit{Sw$_a$}). Assuming an even distribution of flows sourced from \texttt{\small bitcoinminer} (where all flows originate in cooperating networks and traverse the attacked network), each \textit{Sw$_c$} sees 6\,281 outgoing flows, while \textit{Sw$_a$} sees a total of 119\,339 incoming flows, out of which 202 are malicious.
In a centralized setup, each \textit{Sw} sends all observed flows' feature vectors to a \textit{CS}, while the \textit{Sw$_a$} sends its 202 feature vectors. In \sys, \textit{Sw$_a$} sends the 202 feature vectors to the \textit{CM}, which relays them to all \textit{Sw$_c$}. Each \textit{Sw$_c$} performs correlation locally and returns 202 matched flow tuples (192 bits each) to the \textit{CM}.
Table~\ref{tab:bandwidth_summary} shows the communication involved in both scenarios. \sys's distributed design reduces bandwidth usage by 35\%--94.7\%, depending on the sketch. 

\section{Conclusion}

We introduced \sys, a practical framework for distributed attack attribution across cooperating networks. By using compact sketch-based data structures and the orchestration of programmable network elements, \sys is able to accurately correlate malicious flows while maintaining low computational and communication overheads. Our evaluation suggests that flow correlation can be effectively pushed into the network fabric, paving the way for scalable attack attribution.

\bibliographystyle{IEEEtran}
\bibliography{main}

\begin{thebibliography}{10}
\providecommand{\url}[1]{#1}
\csname url@samestyle\endcsname
\providecommand{\newblock}{\relax}
\providecommand{\bibinfo}[2]{#2}
\providecommand{\BIBentrySTDinterwordspacing}{\spaceskip=0pt\relax}
\providecommand{\BIBentryALTinterwordstretchfactor}{4}
\providecommand{\BIBentryALTinterwordspacing}{\spaceskip=\fontdimen2\font plus
\BIBentryALTinterwordstretchfactor\fontdimen3\font minus \fontdimen4\font\relax}
\providecommand{\BIBforeignlanguage}[2]{{%
\expandafter\ifx\csname l@#1\endcsname\relax
\typeout{** WARNING: IEEEtran.bst: No hyphenation pattern has been}%
\typeout{** loaded for the language `#1'. Using the pattern for}%
\typeout{** the default language instead.}%
\else
\language=\csname l@#1\endcsname
\fi
#2}}
\providecommand{\BIBdecl}{\relax}
\BIBdecl

\bibitem{zhang2000detecting}
Y.~Zhang and V.~Paxson, ``Detecting stepping stones.'' in \emph{Proc. of USENIX Security}, vol. 171, 2000.

\bibitem{untangling}
D.~D. Clark and S.~Landau, ``Untangling attribution,'' \emph{Harv. Nat'l Sec. J.}, vol.~2, 2011.

\bibitem{wang2002inter}
X.~Wang, D.~S. Reeves, and S.~F. Wu, ``Inter-packet delay based correlation for tracing encrypted connections through stepping stones,'' in \emph{Proc. of ESORICS}, 2002.

\bibitem{nasr2018deepcorr}
M.~Nasr, A.~Bahramali, and A.~Houmansadr, ``Deepcorr: Strong flow correlation attacks on tor using deep learning,'' in \emph{Proc. of ACM CCS}, 2018.

\bibitem{lopes2024flow}
D.~Lopes, J.-D. Dong, P.~Medeiros, D.~Castro, D.~Barradas, B.~Portela, J.~Vinagre, B.~Ferreira, N.~Christin, and N.~Santos, ``Flow correlation attacks on tor onion service sessions with sliding subset sum,'' in \emph{Proc. of NDSS}, 2024.

\bibitem{murdoch2005low}
S.~J. Murdoch and G.~Danezis, ``Low-cost traffic analysis of tor,'' in \emph{Proc. of IEEE S\&P}, 2005.

\bibitem{palmieri2019distributed}
F.~Palmieri, ``A distributed flow correlation attack to anonymizing overlay networks based on wavelet multi-resolution analysis,'' \emph{IEEE TDSC}, vol.~18, no.~5, 2019.

\bibitem{oh2022deepcoffea}
S.~E. Oh, T.~Yang, N.~Mathews, J.~K. Holland, M.~S. Rahman, N.~Hopper, and M.~Wright, ``Deepcoffea: Improved flow correlation attacks on tor via metric learning and amplification,'' in \emph{Proc. of IEEE S\&P}, 2022.

\bibitem{nasr2017compressive}
M.~Nasr, A.~Houmansadr, and A.~Mazumdar, ``Compressive traffic analysis: A new paradigm for scalable traffic analysis,'' in \emph{Proc. of ACM CCS}, 2017.

\bibitem{sdn-traceback}
Z.~Ling, J.~Luo, D.~Xu, M.~Yang, and X.~Fu, ``Novel and practical sdn-based traceback technique for malicious traffic over anonymous networks,'' in \emph{Proc. of IEEE INFOCOM}, 2019.

\bibitem{coskun2009online}
B.~Coskun and N.~Memon, ``Online sketching of network flows for real-time stepping-stone detection,'' in \emph{Proc. of ACSAC}, 2009.

\bibitem{zheng2023network}
C.~Zheng, X.~Hong, D.~Ding, S.~Vargaftik, Y.~Ben-Itzhak, and N.~Zilberman, ``In-network machine learning using programmable network devices: A survey,'' \emph{IEEE Commun. Surv. Tutor.}, vol.~26, no.~2, 2023.

\bibitem{netbricks}
A.~Panda, S.~Han, K.~Jang, M.~Walls, S.~Ratnasamy, and S.~Shenker, ``Netbricks: Taking the {V} out of $\{$NFV$\}$,'' in \emph{Proc. of USENIX OSDI}, 2016.

\bibitem{scalingHardware}
J.~Sonchack, O.~Michel, A.~Aviv, E.~Keller, and J.~Smith, ``Scaling hardware accelerated network monitoring to concurrent and dynamic queries with *flow,'' in \emph{Proc. of USENIX ATC}, 2018.

\bibitem{mix-match}
L.~Oldenburg, M.~Juarez, E.~A. R{\'u}a, and C.~Diaz, ``Mixmatch: Flow matching for mixnet traffic,'' \emph{PoPETs}, vol. 2024, no.~2, 2024.

\bibitem{holdingAccount}
S.~Staniford-Chen and L.~Heberlein, ``Holding intruders accountable on the internet,'' in \emph{Proc. of IEEE S\&P}, 1995.

\bibitem{zhu2009correlation}
Y.~Zhu, X.~Fu, B.~Graham, R.~Bettati, and W.~Zhao, ``Correlation-based traffic analysis attacks on anonymity networks,'' \emph{IEEE TPDS}, vol.~21, no.~7, 2009.

\bibitem{rf}
M.~Shen, K.~Ji, Z.~Gao, Q.~Li, L.~Zhu, and K.~Xu, ``Subverting website fingerprinting defenses with robust traffic representation,'' in \emph{Proc. of USENIX Security}, 2023.

\bibitem{netwarden}
J.~Xing, Q.~Kang, and A.~Chen, ``Netwarden: Mitigating network covert channels while preserving performance,'' in \emph{Proc. of USENIX Security}, 2020.

\bibitem{barradas2021flowlens}
D.~Barradas, N.~Santos, L.~Rodrigues, S.~Signorello, F.~M. Ramos, and A.~Madeira, ``Flowlens: Enabling efficient flow classification for ml-based network security applications,'' in \emph{Proc. of NDSS}, 2021.

\bibitem{yan_brain-switch_2024}
J.~Yan, H.~Xu, Z.~Liu, Q.~Li, K.~Xu, M.~Xu, and J.~Wu, ``{Brain}-on-{Switch}: {Towards} {Advanced} {Intelligent} {Network} {Data} {Plane} via {NN}-{Driven} {Traffic} {Analysis} at {Line}-{Speed},'' in \emph{Proc. of {USENIX} {NSDI}}, 2024.

\bibitem{defenseSurvey}
X.~Chen, C.~Wu, X.~Liu, Q.~Huang, D.~Zhang, H.~Zhou, Q.~Yang, and M.~K. Khan, ``Empowering network security with programmable switches: A comprehensive survey,'' \emph{IEEE Commun. Surv. Tutor.}, vol.~25, no.~3, 2023.

\bibitem{patronum}
J.~Wu, H.~Pan, P.~Cui, Y.~Huang, J.~Zhou, P.~He, Y.~Li, Z.~Li, and G.~Xie, ``Patronum: In-network volumetric ddos detection and mitigation with programmable switches,'' in \emph{Proc. of ESORICS}, 2024.

\bibitem{superFe}
M.~Zhang, G.~Li, C.~Guo, R.~Yang, S.~Wang, H.~Bao, X.~Li, M.~Xu, T.~Wo, and C.~Hu, ``Superfe: A scalable and flexible feature extractor for ml-based traffic analysis applications,'' in \emph{Proc. of EuroSys}, 2025.

\bibitem{kirci_mass_2022}
E.~C. Kirci, M.~Apostolaki, R.~Meier, A.~Singla, and L.~Vanbever, ``Mass surveillance of {VoIP} calls in the data plane,'' in \emph{Proc. of ACM SOSR}, 2022.

\bibitem{seldom}
E.~Wagner, R.~Matzutt, and M.~Henze, ``Seldom: An anonymity network with selective deanonymization,'' \emph{arXiv preprint arXiv:2412.00990}, 2024.

\bibitem{macedo2022privacy}
I.~Macedo, ``Privacy-preserving machine learning for network traffic analysis,'' MSc. Thesis, Faculdade de Ciências da Universidade do Porto, 2022.

\bibitem{p4ddle}
R.~Doriguzzi-Corin, L.~A.~D. Knob, L.~Mendozzi, D.~Siracusa, and M.~Savi, ``Introducing packet-level analysis in programmable data planes to advance network intrusion detection,'' \emph{Computer Networks}, vol. 239, 2024.

\bibitem{blum2004detection}
A.~Blum, D.~Song, and S.~Venkataraman, ``Detection of interactive stepping stones: Algorithms and confidence bounds,'' in \emph{Proc. of. RAID}, 2004.

\bibitem{fu_detecting_2023}
C.~Fu, Q.~Li, and K.~Xu, ``Detecting {Unknown} {Encrypted} {Malicious} {Traffic} in {Real} {Time} via {Flow} {Interaction} {Graph} {Analysis},'' in \emph{Proc. of NDSS}, 2023.

\bibitem{webb2016finding}
A.~T. Webb and A.~N. Reddy, ``Finding proxy users at the service using anomaly detection,'' in \emph{Proc. of IEEE CNS}, 2016.

\bibitem{nunes2023enhancing}
V.~Nunes, J.~Br{\'a}s, A.~Carvalho, D.~Barradas, K.~Gallagher, and N.~Santos, ``Enhancing the unlinkability of circuit-based anonymous communications with k-funnels,'' \emph{PACMNET}, no. CoNEXT3, 2023.

\end{thebibliography}

\appendix

\section{Appendix}
\label{app}

\begin{figure}[t]
\centering
\begin{subfigure}{0.35\textwidth}
    \centering
    \includegraphics[width=\linewidth]{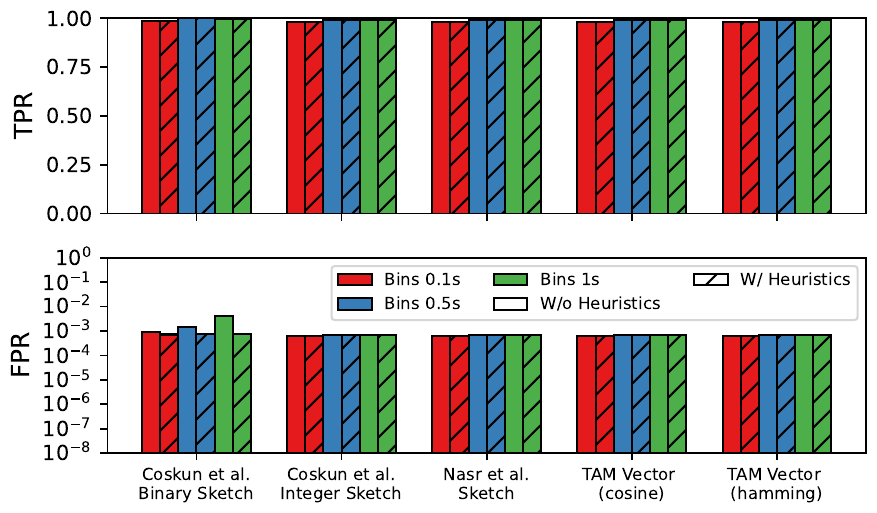}
    \caption{adload}
    \label{fig:adload}
\end{subfigure}

\begin{subfigure}{0.35\textwidth}
    \includegraphics[width=\linewidth]{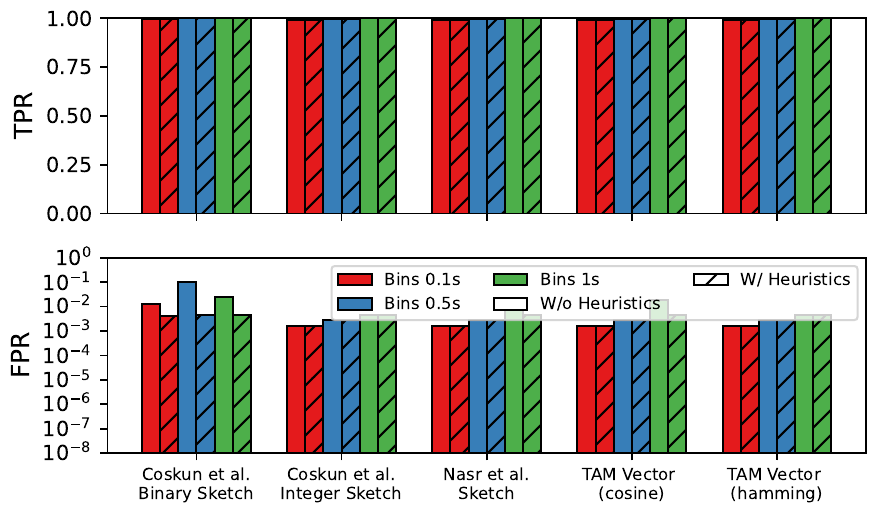}
    \caption{dridex}
    \label{fig:dridex}
\end{subfigure}

\begin{subfigure}{0.35\textwidth}
    \centering
    \includegraphics[width=\linewidth]{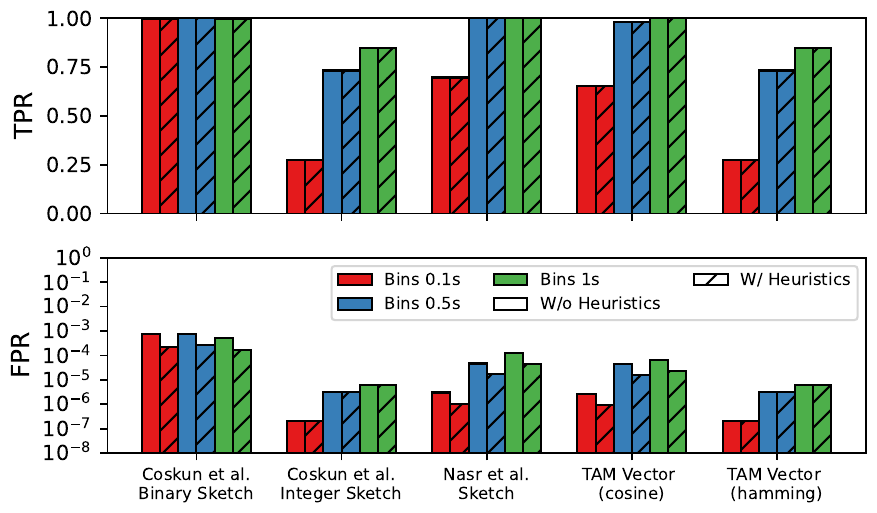}
    \caption{oracle}
    \label{fig:oracle}
\end{subfigure}

\begin{subfigure}{0.35\textwidth}
    \centering
    \includegraphics[width=\linewidth]{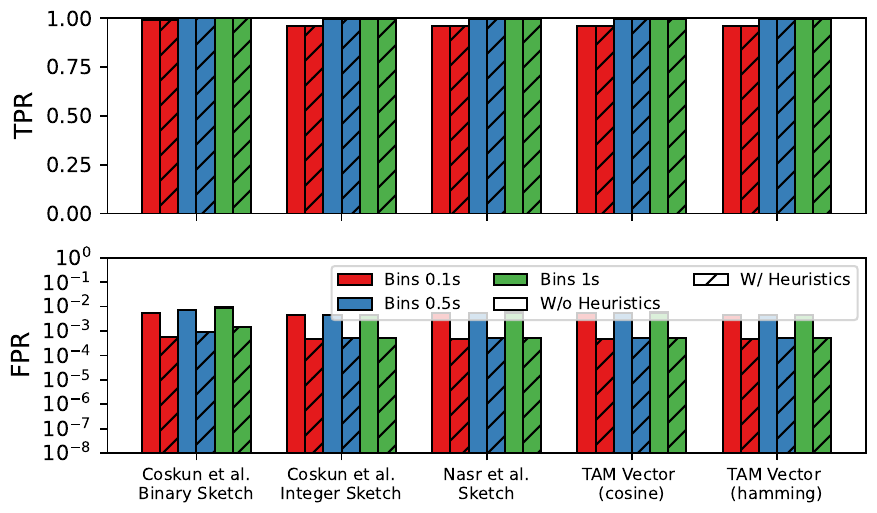}
    \caption{penetho}
    \label{fig:penetho}
\end{subfigure}

\begin{subfigure}{0.35\textwidth}
    \centering
    \includegraphics[width=\linewidth]{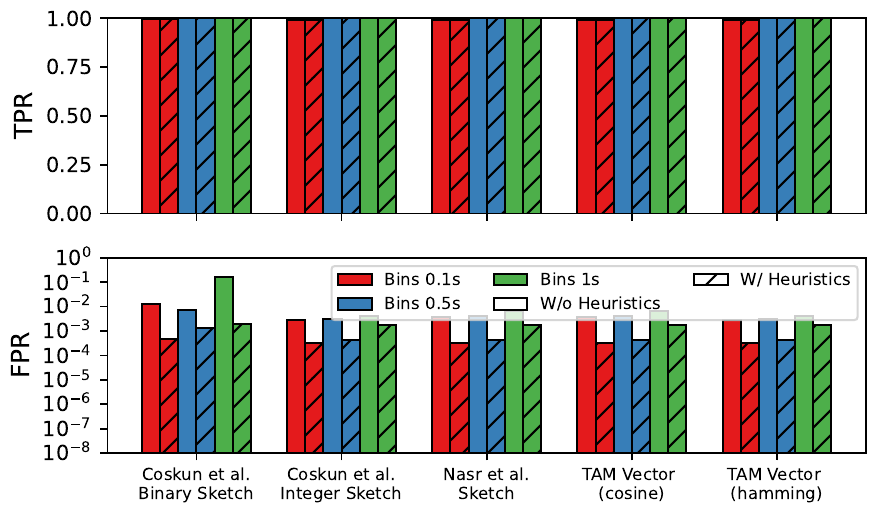}
    \caption{snojan}
    \label{fig:snojan}
\end{subfigure}

\caption{Correlation effectiveness of \sys{} across datasets.}
\label{fig:all_datasets}
\end{figure}

\subsection{Correlation Effectiveness across Multiple Datasets}
\label{app:datasets}

Figures~\ref{fig:adload}--\ref{fig:snojan} show \sys's correlation effectiveness on additional datasets from Fu et al.\cite{fu_detecting_2023}, showing similar trends to those discussed in Section~\ref{sec:eval} on \texttt{\small bitcoinminer}.

The TPR and FPR obtained across all datasets indicate that using the integer sketch by Coskun~\cite{coskun2009online} achieves performance similar to TAM, while requiring 10\% to 60\% less memory (see Figure~\ref{fig:sketch-len}). This suggests that sketching can be effective for reducing memory usage. The results also show that the binary sketch by Coskun et al.~\cite{coskun2009online} achieves higher TPR than TAM in some cases, but with a higher FPR. Both sketches by Nasr et al.~\cite{nasr2017compressive} perform either similarly to or worse than Coskun's sketches. These findings further support the claim made in Section~\ref{sec:eval} that sketches can be used to reduce memory and computation costs while maintaining comparable performance.

\end{document}